\documentclass{emulateapj}
\usepackage{longtable,natbib}
\usepackage[bookmarks=false]{hyperref}

\def\micron{{\mbox{$\mu{\rm m}$}}}

\def\arcmin{{\mbox{$^{\prime}$}}}
\def\degree{{\mbox{$^{\circ}$}}}

\def\C{{\sl Chandra}}

\def\X{{\sl XMM-Newton}}

\begin{document}
\title{Studying Faint Ultra Hard X-ray Emission from AGN in GOALS LIRGs with $Swift$ BAT}

\author{Michael Koss\altaffilmark{1}, Richard Mushotzky\altaffilmark{2}, Wayne Baumgartner\altaffilmark{3}, Sylvain Veilleux\altaffilmark{2,3}, Jack Tueller\altaffilmark{3}, Craig Markwardt\altaffilmark{3}, and Caitlin M. Casey\altaffilmark{1}}
\email{koss@ifa.hawaii.edu}

\altaffiltext{1}{Institute for Astronomy, University of Hawaii, 2680 Woodlawn Drive, Honolulu, HI 96822, USA; }
\altaffiltext{2}{Astronomy Department, University of Maryland, College Park, MD, USA}
\altaffiltext{3}{Astrophysics Science Division, NASA Goddard Space Flight Center, Greenbelt, MD, USA}

\begin{abstract}
We present the first analysis of the all-sky $Swift$ BAT ultra hard X-ray (14-195 keV) data for a targeted list of objects.  We find the BAT data can be studied at 3$\times$ fainter limits than in previous blind detection catalogs based on prior knowledge of source positions and using smaller energy ranges for source detection.  We determine the AGN fraction in 134 nearby (z$<$0.05) luminous infrared galaxies (LIRGS) from the GOALS sample.  We find that LIRGs have a higher detection frequency than galaxies matched in stellar mass and redshift at 14-195 keV and 24-35 keV.  In agreement with work at other wavelengths, the AGN detection fraction increases strongly at high IR luminosity with half of high luminosity LIRGs (50\%, 6/12, $\log L_{IR}/L_{\sun}$$>$$11.8$) detected.   The BAT AGN classification shows 97\% (37/38) agreement with $Chandra$ and $XMM$ AGN classification using hardness ratios or detection of a iron K-alpha line.  This confirms our statistical analysis  and supports the use of the $Swift$ BAT all-sky survey to study fainter populations of any category of sources in the ultra hard X-ray band.  BAT AGN in LIRGs tend to show higher column densities with 40$\pm9$\% showing 14-195 keV/2-10 keV hardness flux ratios suggestive of high or Compton-thick column densities ($\log N_H$$>$$24$ cm$^{-2}$), compared to only 12$\pm$5\% of non-LIRG BAT AGN.  We also find that using specific energy ranges of the BAT detector can yield additional sources over total band detections with 24\% (5/21) of detections in LIRGs at 24-35 keV not detected at 14-195 keV.
\end{abstract}


\section{Introduction}
	The $Swift$ BAT survey with over 500 AGN has revolutionized our study of the ultra hard X-ray sky \citep{Tueller:10:378}, but is still limited to bright ($F_{14-195}$$>$$10^{-11}$~erg s$^{-1}$ cm$^{-2}$) objects in a blind survey. However, the stability of the instrument and the Gaussian nature of the noise, along with its wide energy range, allows the detection at fainter limits for a well defined, moderate sized sample of objects. For the first time, we use this property to study the AGN in luminous infrared galaxies (LIRGs; $\log L_{IR}/L_{\sun}$$>$$11.0$). 

	The nature of the IR (8-1000 $\mu$m) emission and its relation to star formation in AGN is still not well understood.  Past studies of samples of LIRGs have suggested, based primarily on optical and IR AGN indicators,  that the dominant power source is star formation and AGN activity is more common in luminous sources \citep[e.g.][]{Veilleux:95:171}.  Recent studies used a variety of mid-IR spectral diagnostics and X-ray observations \citep{Ptak:03:782,Teng:05:664,Veilleux:09:587, Teng:10:1848, Petric:11:28} to determine the AGN contribution.  However, contamination from star formation and obscuration by dust and gas are problematic.  Additionally, studies of AGN in the hard X-rays have shown the existence of large fraction of  AGN not  showing Spitzer IRAC AGN indicators \citep[e.g.][]{Donley:12:142} and some AGN  are not optically detected \citep{Koss:11:L42,Koss:12:L22}. Since  a significant fraction of X-ray selected AGN are in LIRGs \citep{Koss:11:57},  an ultra-hard X-ray survey of LIRGs might come to different conclusions than those derived at lower energies. 
	  	
	The ultra hard X-rays ($>$15 keV) are much less sensitive to obscuration in the line-of-sight than soft X-ray or optical wavelengths and are biased only against highly Compton-thick AGN \citep{Burlon:11:58}.  This band is also free from contamination from star formation that is significant in the soft  X-rays ($<$5 keV).  Additionally, in Compton-Thick AGN the radiation below 10 keV is almost completely absorbed in the X-rays whereas a broad Compton reflection hump appears in the $>$15 keV continuum (Reynolds 1998).  Thus, ultra-hard X-ray observations are an important complement to lower energy X-ray data. 

	
	We use the the most sensitive all-sky ultra hard X-ray survey from the $Swift$ BAT instrument to search for AGN emission in LIRGs.  Previous studies using the $INTEGRAL$ satellite  stacked emission from a large sample of IRAS bright galaxies and found no AGN detection \citep{Walter:09:97}.  Additionally, past BAT AGN catalogs generated $>$$4.8\sigma$ sources from "blind" detections  \citep[e.g.][]{Tueller:10:378}.  To achieve higher sensitivities, we identify AGN (see $\S$2.2) based on the prior knowledge of source positions and search in energy bands where we expect the AGN emission to be brightest.    We adopt a standard cosmology ($\Omega_m$=0.3, $\Omega_\Lambda$=0.7, $H_0$ =70 km s$^{-1}$ Mpc$^{-1}$) to determine distances. 
\section{Sample Selection and Derived Quantities}

\subsection{Sample of LIRGs and ULIRGs}
	We selected a sample of nearby LIRGs (z$<$$0.05$) in the northern hemisphere (DEC$>$-25) from the Great Observatories All Sky LIRG Survey \citep[GOALS;][]{Armus:09:559}.  In this redshift range, we are sensitive to X-ray luminosities of $L_{14-195 \; keV}$$>$10$^{42.0}$ erg/s.  This limit effectively detects AGN since it is ten times larger than the maximum known emission from a starburst galaxy  (e.g., M82, $\log L_{14-195 \; keV}$$=$40.8 erg/s).   Since single temperatures and SED templates can overestimate IR luminosities  \citep{Casey:12:1595}, we recomputed GOALS IR luminosities based on SED fitting using IRAS data and a model joining a modified, single dust temperature greybody, that approximates hot-dust emission from AGN heating.  We have limited our sources to be outside the Galactic plane (b$>$10$\degree$) because of source confusion in IRAS and Swift, as well difficulty measuring stellar masses because of high levels of optical extinction.    
	
	We have also limited our sample because in the low resolution BAT detector, source confusion from nearby bright AGN can occur.  For blind source detection, \citet{Ajello:09:367} estimated a confusion radius of 5.5$\arcmin$ at SNR=2, 3.8$\arcmin$ at SNR=3,  and 2.8$\arcmin$ at SNR=4.  We use a conservative approach and exclude all detections within 15$\arcmin$ of BAT catalog sources.   This excludes seven LIRGs from our study.  NGC 232 and NGC 838 are in merging galaxy groups with a nearby ($<$$2\arcmin$) bright BAT-detected AGN companion.    Additionally, UGC 3608 is 5.1$\arcmin$ from a nearby ROSAT X-ray source, 1RXS J065711.8+462731.  A 1.5 ks XRT observation suggests the majority of the flux is coincident with this ROSAT source.  IRAS F03217+4022 and UGC 02717 are near a bright BAT AGN, IRAS 03219+4031 at 8.5$\arcmin$ and 7$\arcmin$ separation respectively.  Finally, NGC 2524 is near a bright BAT AGN Mrk 0622 at 10.7$\arcmin$.   This leaves our total LIRG sample with 134 objects.  


\subsection{Faint BAT Source Detection in the GOALS Sample}

	In $Swift$ BAT, the detector noise distribution is a Gaussian function centered at zero significance.   Real astrophysical sources show a tail in the distribution at positive significances.  Significant detections in the blind BAT detection catalogs are defined at $>$4.8 SNR to ensure zero false sources caused by random fluctuations in a large sample ($\approx$500).  Source detection is performed on a map weighted to the Crab Nebula, using a single average map of all eight energy bins between 14-195 keV.   
	
	However, many real astrophysical sources are below 4.8 SNR and can be studied based on known positions of galaxies and by studying energy range where the source population is brightest.  Using the 24-35 keV energy bin for instance, we are more sensitive to the reflection component of obscured AGN.  We use the $BATCELLDETECT$ software which performs a sliding cell method to locate regions of the image which are significantly different from the background.  We simultaneously fit all of the 1092 previously detected BAT AGN in the 70 month catalog along with the LIRGs in the 14-195 keV band and 24-35 keV band.  Additionally, we select a comparison sample of 1000 galaxies matched in stellar mass and redshift from the NASA-Sloan Atlas \citep{Blanton:11:31}. To compute the stellar mass of the LIRGs and galaxy control sample, we use \textit{ugriz} photometry following \citet{Koss:11:57} using the software $kcorrect$ v4.2 and SDSS imaging.  For galaxies in close mergers, we follow \citet{Koss:10:L125} and estimate the stellar mass from the largest galaxy.

	
	We use the distribution of SNR for 1000 random pointings from the SDSS survey area to measure the significance of the X-ray detections in the other samples.  The significance distribution at 14-195 keV of the random pointings is well fit by a Gaussian centered at 0.02$\pm$0.11 SNR with $\sigma$=1.01$\pm$0.06, consistent with the expected values for a Gaussian distribution of pure noise.  For the LIRG and matched sample, the Gaussian distribution of noise is fit from the SNR$<$0 source distribution.  
	
	We choose a 2.7$\sigma$ cutoff SNR in the 134 LIRG sample to have on average less than one 'fake' noise source based on the Gaussian distribution of noise using both whole band 14-195 keV and 24-35 keV detections if we assume the distribution is pure noise.  Finally, we note that the lowest SNR of any LIRG is -2.1 at 14-195 keV and -2.7 at 24-35 keV suggesting this cutoff should assure a sample of clean individual detections. 
	
	
	We also analyze X-ray emission using XSPEC v12.7.1 for the new sources between 2.7-4.8$\sigma$.  To calculate luminosities and upper limits, we assume an X-ray power law of $\Gamma$$=$1.9 and Galactic extinction, consistent with the mean 14-195 keV power law for Seyfert 2s in the 70 month blind detection catalog \citep{Winter:09:1322}. The BAT emission is absorbed by $<$10\% for $N_H$$<$$3\times10^{23}$ cm$^{-2}$, but sources with larger obscurations are underestimated.   To determine 1$\sigma$ errors in luminosity, we include the error from assuming a fixed power law index ($14\%$) as well as sky and detector noise ($<$37\%).   Finally, to better understand the average properties of the sources, we fit a simple X-ray power law to the average emission in each X-ray band.

\subsection{X-ray Hardness Flux Ratios and Comparison Sample} 
 	The ultra hard X-ray hardness flux ratio ($HR_{UX}$=14-195 keV/2-10 keV) provides a measure of obscuration in heavily obscured AGN ($N_H$$>$$10^{23}$ cm$^{-2}$) since the transmitted hard X-ray emission is suppressed by a much larger factor than the ultra hard X-ray emission.  Long term AGN variability can affect this ratios, but this variability is typically 20-40\% in the 2-10 keV X-rays \citep{McHardy:01:205} and smaller in the ultra hard X-rays \citep{Ricci:11:102}.   
	
	To estimate absorbing columns corresponding to $HR_{UX}$,  we use the MYTorus model \citep{Murphy:09:1549}, which fully treats photoelectric absorption and relativistic Compton scattering.  The intrinsic AGN emission was modeled as a power law ($\Gamma$=1.9)  and the column density assumes the torus is seen edge-on following \citet{Burlon:11:58}.  In this model, the emission is reduced by four at $N_H$$=$$3\times10^{23}$ cm$^{-2}$ and $N_H$$=$$4\times10^{24}$ cm$^{-2}$ for 2-10 keV and 14-195 keV, respectively, showing that the ultra hard X-rays can pass through an order of magnitude higher absorbing column.    
	 	 
	 Finally, as a comparison sample we measured the $HR_{UX}$ from 49 non-LIRG BAT-detected AGN ($\log L_{IR}/L_{\sun}$$<$$11.0$) from \citet{Winter:09:1322} from the same redshift range to understand whether BAT AGN in LIRGs have higher levels of obscuration.



\clearpage

\begin{figure} 
\includegraphics[width=8.8cm]{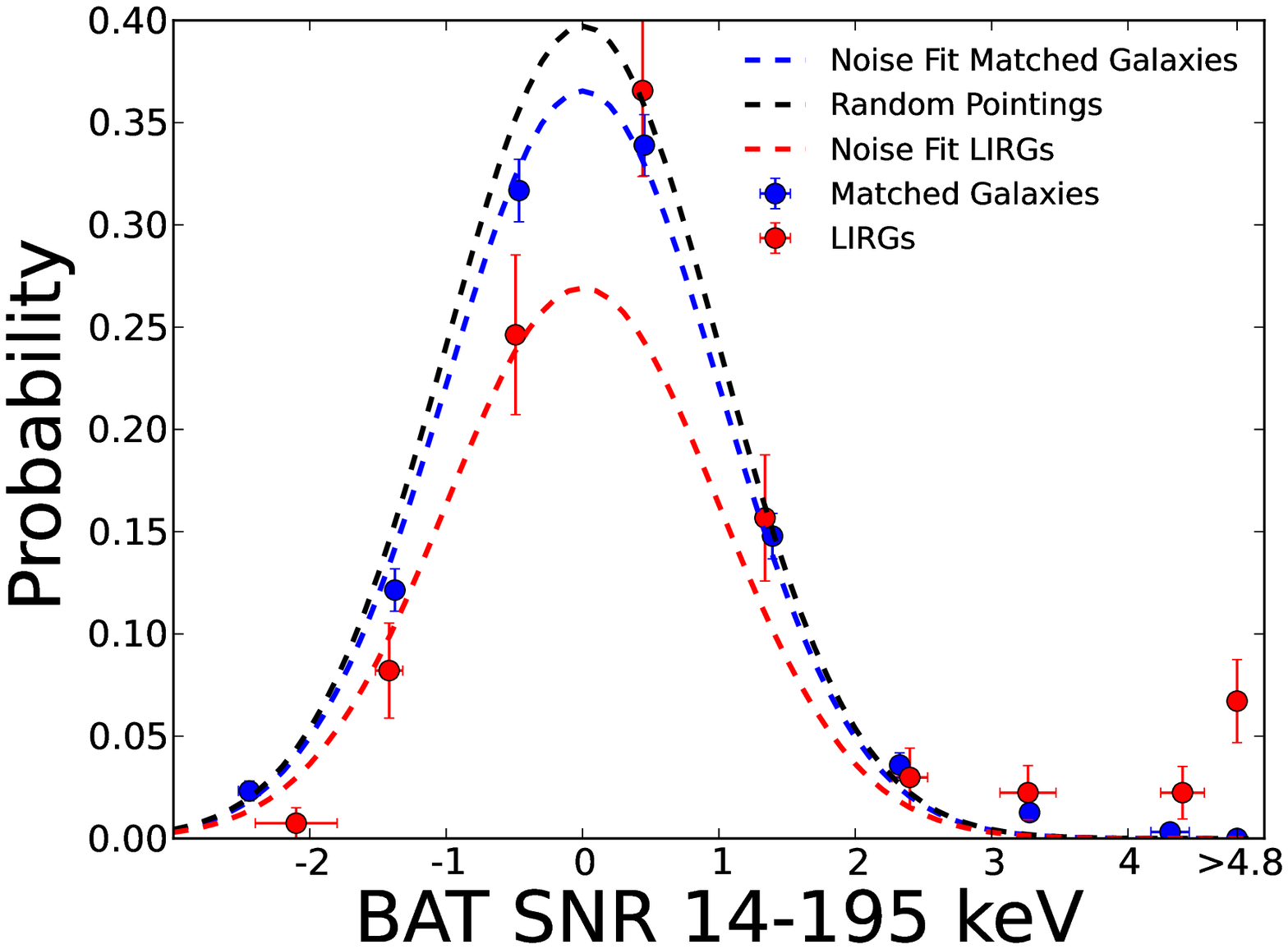}
\includegraphics[width=8.8cm]{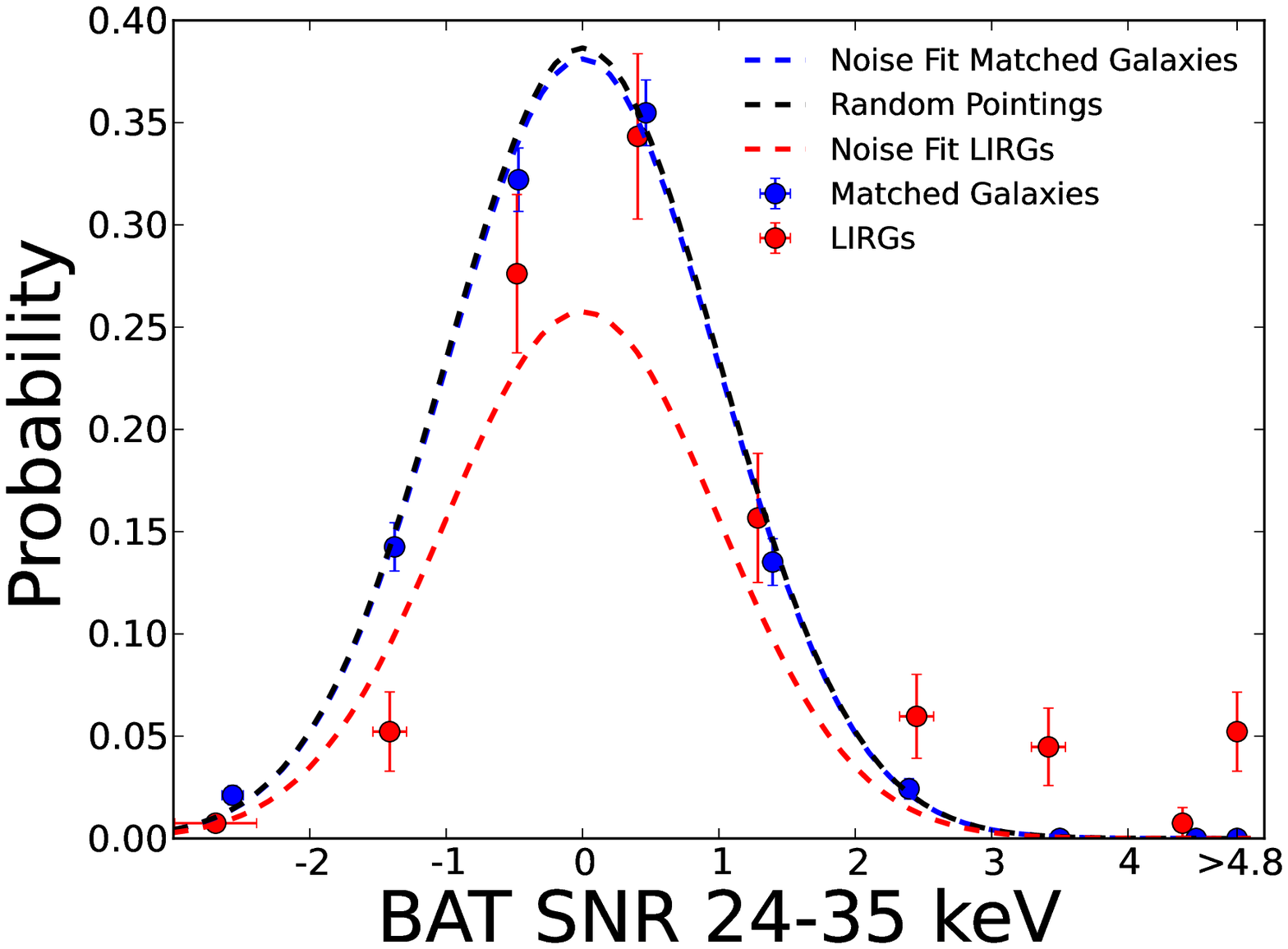}
\caption{Histogram of mean BAT signal to noise ratios (SNR) for the LIRGs and a sample matched in stellar mass and redshift for 14-195 keV (left) and 24-35 keV (right).  The dashed lines indicate the best fit lines for the expected Gaussian noise in each sample.   The black dashed lines indicate the expected distribution from random pointings.  Error bars are computed from 1,000 bootstrapping trials.  The LIRG sample has a higher detection frequency than the matched galaxy sample and random pointings.}
\label{leddmorph}
\end{figure}  

\section{Results}

\subsection{The Fraction of AGN in U/LIRGs}
		
	We compare the BAT detection significance at 14-195 keV and 24-35 keV of the LIRGs and a sample of galaxies matched in stellar mass and redshift (Fig.~1).   There is an excess of LIRGs at SNR$>$3 for 14-195 keV and SNR$>$2 at 24-35 keV,  based on the distributions from random pointings.  There is an excess of LIRG detections over the matched galaxies at SNR$>$4 at 14-195 keV and at SNR$>$2 at 24-35 keV.  The fraction in the LIRG sample above 2.7$\sigma$ at 14-195 keV is 11\%$\pm$2 (16/134) and at 24-35 keV 14\%$\pm$2 (19/134), while the matched galaxy sample is only 2$\pm$1\% (21/1000) at 14-195 keV and 1$\pm$1\% (5/1000) at 24-35 keV.   This suggests that LIRGs are more likely to be detected as ultra hard X-ray AGN than galaxies of a similar stellar mass and redshift consistent with previous results \citep[e.g.][]{Koss:11:57}.  Although we cannot reliably identify individual sources below 2.7 SNR, analysis of Gaussian fits to the negative SNR distribution representative of noise, show a total LIRG sample detection fraction of 33$\pm8$\% at 14-195 keV and  36$\pm7$\% at 24-35 keV compared to the matched galaxy sample detection fraction of only 9$\pm2$\% at 14-195 keV and 4$\pm1$\% at 24-35 keV.
	
	We also analyze the average emission in each BAT energy band (Fig.~2).  We find there is a significant excess among stacked sources at $>$1 SNR, between energies of 14-150 keV (Fig.~2-left).     For the $>$4.8 SNR sources, we find they are fit by a power law with index $\Gamma$=2.18$\pm$0.30 and for the sources above the cutoff (2.7$<$SNR$<$4.8), we find a harder spectrum of $\Gamma$=1.51$\pm$0.22 (Fig.~2-right).   
	
	We also look for additional sources detected in the 24-35 keV band where the reflection component of Compton Thick AGN is expected to contribute significantly that are not detected in the 14-195 keV band.  Of the 12 new SNR=2.7-4.8 detections, five are detected in the 14-195 keV band with a stronger significance than the 24-35 keV, and the remaining 7 are detected with stronger significances in the 24-35 keV band.   More than half (12/21, 57\%) of detected LIRGs are at SNR=2.7-4.8 and thus not detected in previous BAT catalogs.    
	
	The BAT detection fraction  of LIRGs at 24-35 and 14-195 keV is shown in Figure 3.  The AGN detection fraction increases strongly at high IR luminosity ($\log L_{IR}/L_{\sun}$$>$11.8) with half 50\% (6/12) detected above 2.7 SNR.   
	
\begin{figure} 
\includegraphics[width=8.cm]{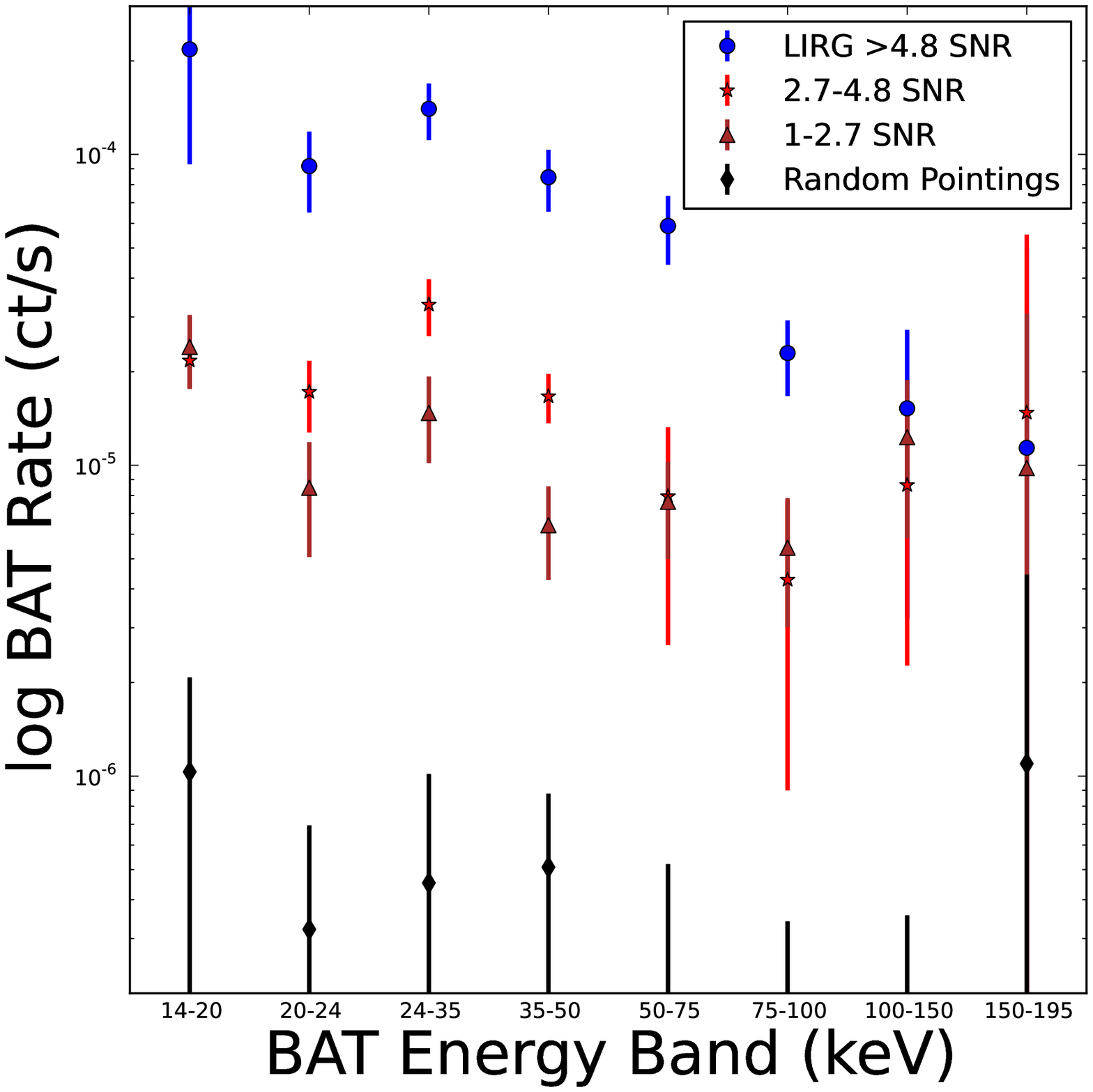}
\includegraphics[width=7.8cm]{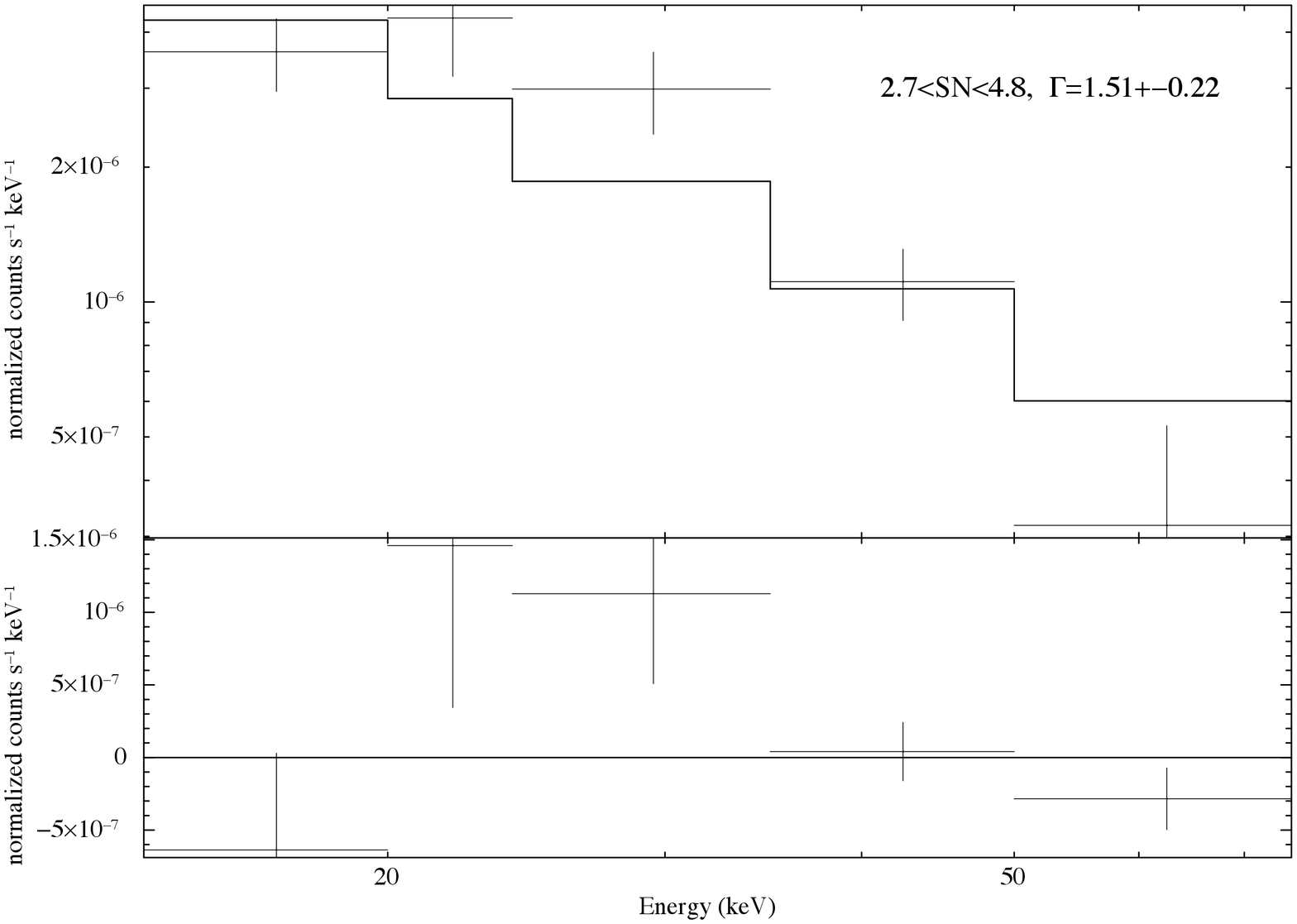}
\caption{{\em Left}:  Stacked count rates in each of the 8 BAT energy bands.  We find an excess of emission for sources above 1$\sigma$ for energies 14-150 keV.  {\em Right}:  Stacked spectra for sources between 2.7-4.8$\sigma$ with residuals shown.  A fit with a simple power law has a best fit of $\Gamma$=1.51$\pm$0.22, consistent with the power law of other BAT sources.  Residual emission is found at 20-35 keV, consistent with fact that 8/9 of the sources with high quality X-ray data are Compton Thick AGN.}
\label{leddmorph}
\end{figure}  

\subsection{Comparison with 2-10 keV Classification}
	We compare the AGN classification using  $\C$ and $\X$ of previous LIRG samples based on hardness ratios of the X-ray spectra (HR$>$-0.3) or the detection of an Fe K$\alpha$ line with the BAT classification.  The C-GOALS $\C$ survey \citep{Iwasawa:11:106} classified luminous LIRGs ($L_{IR}$$>$11.73). The BAT and $\C$ classifications agree for 18/19 galaxies common in both samples.	VV 340a, a Compton-thick AGN is detected in C-GOALS, but not in BAT (SNR=0.57).  More nearby studies of less luminous LIRGs were done by \citet{Lehmer:10:559} and \citet{Pereira-Santaella:11:93} using $\C$ and $\X$.  We find agreement with BAT in 20/20 cases in these samples.  Overall we find agreement in 38/39 cases or 97\% based on hardness ratios or the detection of an Fe K$\alpha$ line.  There are other cases where a lower luminosity AGN is detected using the ratio of the galaxy nucleus to total galaxy emission in the 2-8 keV band \citep[NGC 4194, NGC 7771;][]{Lehmer:10:559} that are not detected as AGN in BAT. 
	
\begin{figure} 
\includegraphics[width=8.8cm]{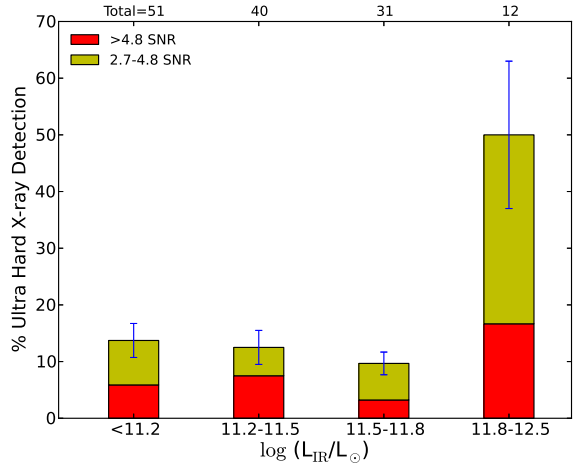}
\caption{{\em Left}:  GOALS LIRGs IR luminosity by BAT SNR.  We find a higher fraction of detections at the highest IR luminosity.  Error bars are based on statistical error assuming poisson statistics.  The trend is in agreement with previous emission line diagnostics work \citep[e.g.][]{Veilleux:95:171}. }
\label{leddmorph}
\end{figure}

	

	We also compare the AGN classification for 12 new SNR=2.7-4.8 detections.  Five of these sources are detected in the 14-195 keV band with a stronger significance than the 24-35 keV, and 7 are detected with stronger significances in the 24-35 keV band.  NGC 7674, UGC 5101, NGC 6926, UGC 08696 (Mrk 273), UGC 08058 (Mrk 231), NGC 3690 are Compton Thick \citep{Severgnini:12:46}   IRAS F17207-0014 shows the presence of strong (at 2$\sigma$), high-ionization Fe K line on a hard continuum \citet{Iwasawa:11:106}.  UGC 2608 is also listed as a heavily obscured Compton-Thick AGN \citep[$N_H$$>$$10^{24}$ cm$^{-2}$,][]{Guainazzi:05:119}.  A Chandra observation of NGC 1961 has a hardness ratio indicative of an AGN (HR=-0.2).  Mrk 331 has a hardness flux ratio indicative of star formation (HR=-0.5), no significant Fe K line, weak 2-10 keV emission ($\log L_{2-10 \; keV}$$=$40.7), however it does have a compact radio source suggesting an AGN \citet{Parra:10:555}.  IRAS F02437+2122 has no high quality X-ray data, but is a LINER AGN \citep{Veilleux:95:171}.   UGC 3094 has a Ne V detection suggesting the presence of an AGN  \citet{Petric:11:28}.  Finally, NGC 0877 has no high quality observation to test for the presence of an AGN.

\subsection{Comparison with Spitzer AGN Classification}	
	The Ne V lines at 14.3 and 24.3 $\micron$ imply the presence of an AGN since this line requires 97 eV and is too large to be produced even by O stars.  There are 29 LIRGS with Ne V detections overlapping in our sample with the \citet{Petric:11:28} Spitzer study (Table 1), with 14/29 (48\%) detected in BAT.  Conversely, 14/21 (67\%) of BAT-detected LIRGs have Ne V.  The 33\% non-detection in Ne V for BAT-detected LIRGs is lower than the 10\% found by \citet{Weaver:10:1151} for all BAT AGN.  However, this study use a deeper exposure map which is more sensitive to fainter sources  (70 vs.~9 months), as well as fainter detection limits (2.7$<$SNR$<$4.8), and is exclusively of LIRGs which may be more likely to have optically thick, dusty gas close to the AGNs \citep[e.g.,][]{Armus:07:148}.  We note that 5/6 of the sources without Ne V detections have X-ray, optical, or radio observations confirming the presence of AGN (see \S3.2).  
	
\subsection{Properties of AGN in LIRGs Compared to non-LIRGs}
	 
	Previous hard  X-ray observations have found some LIRGs to be heavily obscured Compton-thick AGN \citep[NGC 6240, NGC 3690, UGC 5101]{Komossa:08:86,dellaCeca:02:L9,Imanishi:03:L167}.  We measure $HR_{UX}$ to test whether LIRGs  are more obscured than non-LIRG BAT AGN (Fig.~4-right).  The median HR$_{UX}$=15 among LIRGs corresponds to a $N_H$$\approx$$4\times10^{23}$ cm$^{-2}$ compared to a median HR$_{UX}$=3.8 or $N_H$$\approx$$7\times10^{22}$ cm$^{-2}$ for non-LIRG BAT AGN.  A Kolmogorov-Smirnov (K-S) test indicates a ($<$1\%) chance that the HR$_{UX}$ from the samples are from the same distribution indicating that LIRGS show systematically higher column densities.
	
\begin{figure} 

\includegraphics[width=6.1cm]{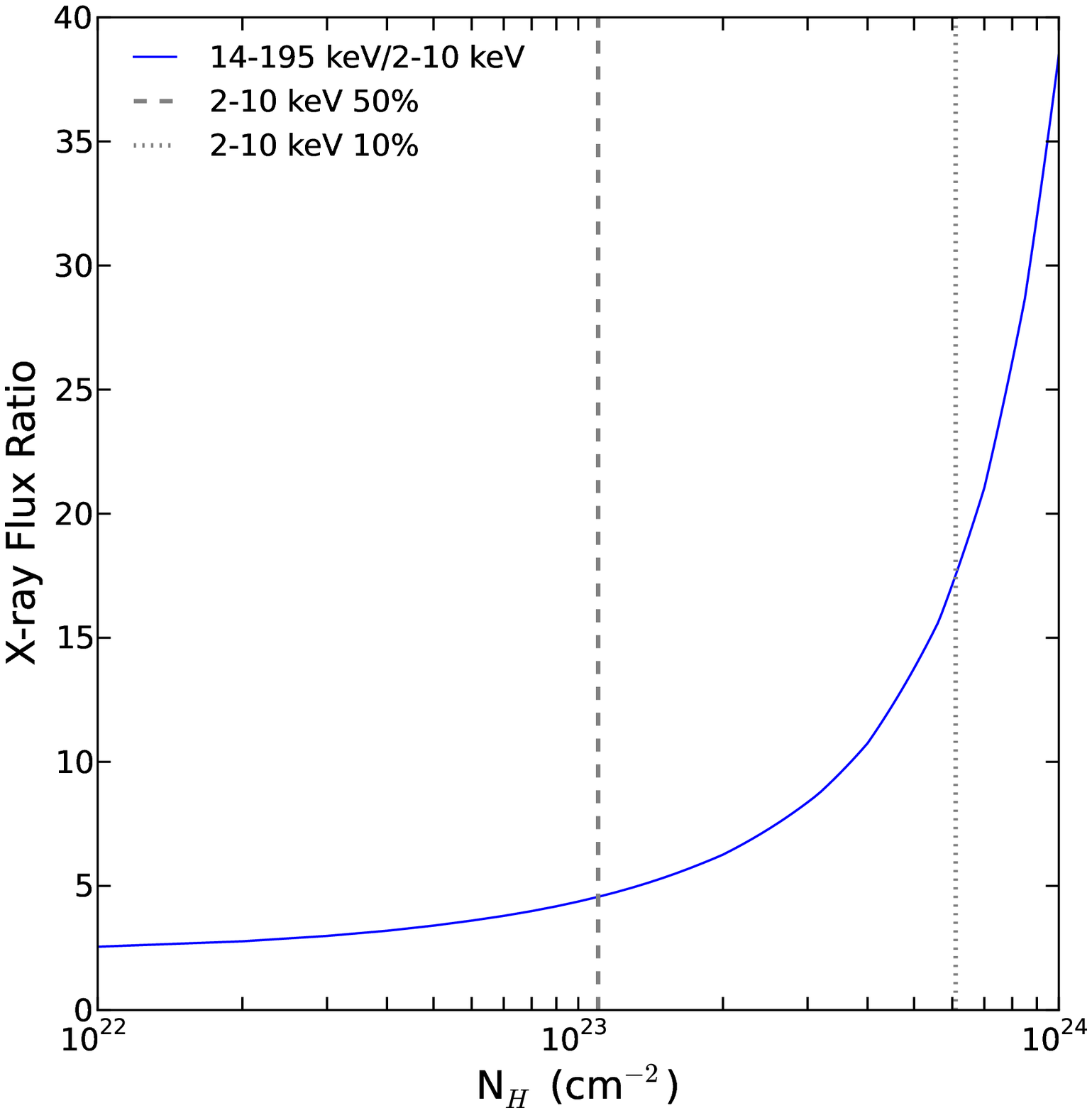}
\includegraphics[width=8.8cm]{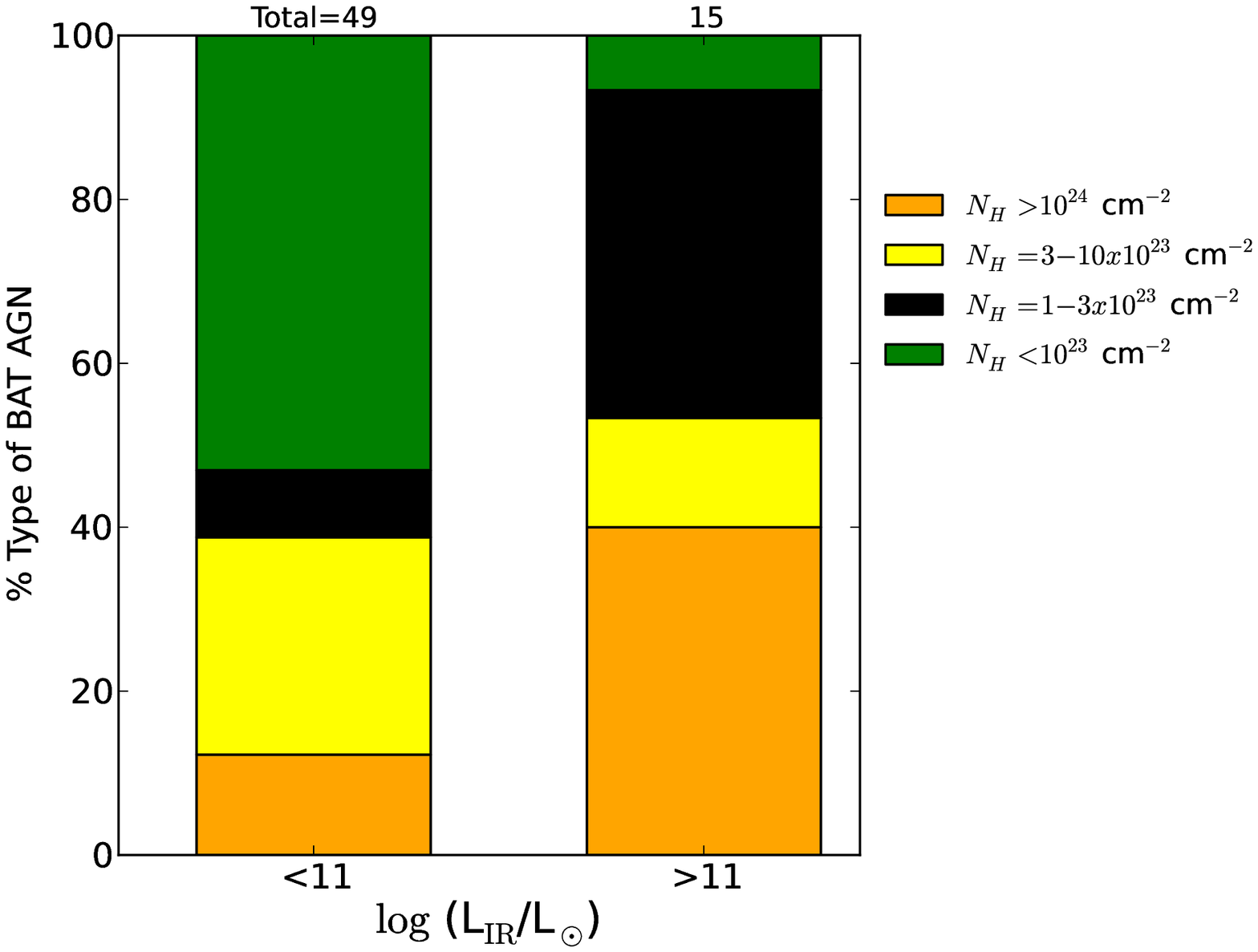}
\caption{{\em Left}:  Ultra hard X-ray hardness flux ratio ($HR_{UX}$=14-195 keV/2-10 keV) for an AGN with a power-law spectrum index of 1.9, as a function of the column density of the torus as seen edge-on.  This measure provides a measure of obscuration since the transmitted hard X-ray emission is suppressed by a much larger factor than the ultra hard X-ray emission.  {\em Right}:  Approximate column density from $HR_{UX}$  as a function of IR luminosity.  BAT-selected AGN in LIRGs tend to show higher column densities than non-LIRG AGN.}
\label{leddmorph}
\end{figure}  
	



\section{Summary and Discussion}	  

We search for nuclear activity in nearby LIRGs based on the detection of ultra high X-ray emission from Swift BAT. We find:

\begin{enumerate}
\renewcommand{\theenumi}{(\roman{enumi})}
\renewcommand{\labelenumi}{\theenumi}

\item A lower cutoff (SNR$>$2.7) than previous 'blind' catalogs (SNR$>$4.8) can be used for a moderate sample size ($\approx$100).  Using this cutoff at 14-195 keV and 24-35 keV, we find agreement in AGN classification for 38/39 cases (97\%) from Chandra and XMM  based on hardness ratios or the detection of an Fe K$\alpha$ line.

\item  We find that using specific energy ranges of the BAT detector can yield additional sources over single band detections with 24\% (5/21) of detections in LIRGs at 24-35 keV not detected at 14-195 keV.  Of the 12 new SNR=2.7-4.8 detections, 7 are detected with stronger significances in the 24-35 keV band than the 14-195 keV band. 

\item LIRGs have a higher BAT-detection frequency at 14-195 keV and 24-35 keV compared to galaxies matched in stellar mass.  Additionally, the BAT-detection fraction increases strongly at high IR luminosities with half of high luminosity LIRGs detected (50\%, 6/12, $\log L_{IR}/L_{\sun}$$>$11.8).

\item BAT detected AGN in LIRGs have higher column densities with 40$\pm9$\% (6/15) having $HR_{UX}$ suggestive of high column densities ($\log N_H$$>$24 cm$^{-2}$), compared to only 12$\pm$5\% (6/49) of non-LIRG BAT AGN.  Additionally, 8/9 of the new SNR=2.7-4.8 BAT sources with high quality X-ray data are Compton-Thick based on past observations.  We also find the stack spectra of these new sources show an excess at 24-35 keV consistent with a reflection component in a Compton-thick AGN.




\end{enumerate}

	
	We note that there are several LIRGs in warm infrared sources detected in the ultra hard X-rays (i.e. UGC 07064, MCG +08-11-011, NGC 5995, Mrk 520, NGC 1142, Mrk 463) that are not included in the GOALS sample because of the 60 $\micron$ cutoff, therefore this study underestimate the total fraction of AGN in all LIRGs based on the ultra hard X-rays.  These sources are predominantly unobscured Seyfert 1s where the AGN contributes significantly to the total IR emission and will be discussed in a forthcoming paper.  

	These results show the potential to use the $Swift$ BAT all-sky survey to study $\approx$3$\times$ fainter populations of ultra hard X-ray sources than the past catalogs based on source positions and by using certain energy ranges where the sources are expected to be brightest.  A different survey could study faint BAT-detection in obscured AGN, star forming galaxies, radio loud AGN, or galactic sources.  Additionally, since lower energy all-sky surveys such as ROSAT show little or no correlation in count rates with $Swift$ because of the effects of obscuration \citep{Markwardt:05:L77}, this remains an important all-sky resource to utilize with small field of view X-ray missions.   For instance, this technique could be used to identify promising candidates to study with higher sensitivity and resolution small field of view missions such as $NuSTAR$ and $Astro$-$H$.  
	
	  The success of $Swift$ in identifying similar numbers of AGN in nearby LIRGs (z$<$0.05) as $Chandra$ and $XMM$ suggests that higher sensitivity missions such as $NuSTAR$ and $Astro$-$H$ hold great promise to study even more distant, obscured AGN (z$>$0.05, $N_H$$>$$10^{24}$ cm$^{-2}$) since they can reach these all-sky sensitivities in only 15 minutes. 
	  

	
\section{Acknowledgements}
We acknowledge the $Swift$ BAT team and are grateful to Ezequiel Treister and Marco Ajello for discussion and suggestions.





\begin{center}
\begin{longtable}{l c c c c c c c} 
\caption{\textbf{Properties of GOALS LIRGs}}  \label{f33} \\
\hline \hline \\[-2ex]
   \multicolumn{1}{c}{Galaxy} &
   \multicolumn{1}{c}{$L_{IR}$\tablenotemark{1}} &
   \multicolumn{1}{c}{SNR\tablenotemark{2}} &
   \multicolumn{1}{c}{SNR} &
   \multicolumn{1}{c}{$L_{14-195\,keV}$\tablenotemark{3}} &
   \multicolumn{1}{c}{HR$_{UX}$\tablenotemark{4}} &
   \multicolumn{1}{c}{NeV\tablenotemark{5}} &
   \multicolumn{1}{c}{X-ray\tablenotemark{6} } \\
   
   \multicolumn{1}{c}{Name} &
   \multicolumn{1}{c}{$\log L_{\sun}$} &
   \multicolumn{1}{c}{14-195 keV} &
   \multicolumn{1}{c}{24-35 keV} &
   \multicolumn{1}{c}{$\log$ erg/s} &
   \multicolumn{1}{c}{Diag} &
   \multicolumn{1}{c}{X-ray} &
   \multicolumn{1}{c}{Ref} 

\\[0.5ex] \hline
   \\[-1.8ex]
\endfirsthead

\multicolumn{3}{c}{{\tablename} \thetable{} -- Continued} \\[0.5ex]
   \multicolumn{1}{c}{Galaxy} &
   \multicolumn{1}{c}{$L_{IR}$\tablenotemark{1}} &
   \multicolumn{1}{c}{SNR\tablenotemark{2}} &
   \multicolumn{1}{c}{SNR} &
   \multicolumn{1}{c}{$L_{14-195\,keV}$\tablenotemark{3}} &
   \multicolumn{1}{c}{HR$_{UX}$\tablenotemark{4}} &
   \multicolumn{1}{c}{NeV\tablenotemark{5}} &
   \multicolumn{1}{c}{X-ray\tablenotemark{6} } \\
   
   \multicolumn{1}{c}{Name} &
   \multicolumn{1}{c}{$\log L_{\sun}$} &
   \multicolumn{1}{c}{14-195 keV} &
   \multicolumn{1}{c}{24-35 keV} &
   \multicolumn{1}{c}{$\log$ erg/s} &
   \multicolumn{1}{c}{Diag} &
   \multicolumn{1}{c}{X-ray} &
   \multicolumn{1}{c}{Ref} 
   
       \\[0.5ex] \hline
   \\[-1.8ex]
\endhead

{\em \textbf{BAT Detections}}\\
\hline
IRAS F02437+2122& 11.13& 0.7& 2.7& $<$42.9&?&&\\
IRAS F05189-2524& 12.14& 6.0& 4.2& 43.72$_{-0.09}^{+0.12}$&4&Y&I11\\
IRAS F17207-0014& 12.36& 1.1& 3.3& $<$43.6&?&&\\
MCG-03-34-064& 11.13& 12.8& 7.3& 43.25$_{-0.05}^{+0.05}$&15&Y&W09\\
MCG+04-48-002& 10.84& 26.8& 16.4& 43.54$_{-0.02}^{+0.02}$&26&Y&W09\\
MRK 0331& 11.42& 1.9& 2.7& $<$42.6&?&&\\
NGC 0877& 10.99& 2.3& 3.2& $<$42.4&?&&\\
NGC 1068& 11.37& 15.6& 11.6& 42.03$_{-0.04}^{+0.04}$&99&Y&L10\\
NGC 1275& 11.25& 50.2& 19.5& 43.69$_{-0.01}^{+0.01}$&6&&P06\\
NGC 1961& 10.87& 2.3& 2.9& $<$42.4&?&&\\
NGC 3690& 11.76& 2.9& 3.0& 42.04$_{-0.17}^{+0.27}$&10&&I11\\
NGC 6240& 11.81& 18.8& 13.8& 43.96$_{-0.03}^{+0.03}$&29&Y&I11\\
NGC 6926& 11.17& 2.8& 3.6& 42.81$_{-0.18}^{+0.3}$&126&Y&G08\\
NGC 7469& 11.57& 28.5& 16.6& 43.61$_{-0.02}^{+0.02}$&2&Y&W09\\
NGC 7674& 11.56& 4.2& 2.7& 43.28$_{-0.12}^{+0.18}$&20&Y&L10\\
NGC 7679& 11.05& 6.5& 3.8& 43.01$_{-0.08}^{+0.11}$&5&Y&P11\\
PGC 016795& 11.20& 9.3& 6.3& 43.25$_{-0.06}^{+0.07}$&38&Y&XRT\\
UGC 03094& 11.38& 3.1& 1.6& 43.05$_{-0.16}^{+0.26}$&?&Y&\\
UGC 05101& 11.96& 4.8& 3.6& 43.43$_{-0.11}^{+0.15}$&53&Y&I11\\
UGC 08058 (Mrk 231)& 12.50& 3.7& 1.5& 43.34$_{-0.14}^{+0.21}$&7&Y&I11\\
UGC 08696 (Mrk 273)& 12.11& 4.4& 2.4& 43.31$_{-0.12}^{+0.17}$&8&Y&I11\\
\hline
{\em \textbf{BAT Non-Detections}}\\
\hline
IRAS 04271+3849& 11.14& 2.2& 0.5& $<$42.9&&Y&\\
NGC 5256& 11.47& 1.9& 2.2& $<$42.9&&&\\
NGC 4418& 10.99& 1.8& 1.3& $<$43.2&&&\\
IRAS 18090+0130& 11.56& 1.8& -0.5& $<$43.2&&&\\
IC 5298& 11.57& 1.7& -0.1& $<$43.0&&&\\
MCG -02-33-098& 11.01& 1.5& 1.0& $<$42.6&&&\\
NGC 5104& 11.10& 1.5& 0.4& $<$42.7&&&\\
ESO 557-G002& 11.11& 1.4& 1.5& $<$42.8&&&\\
NGC 5990& 10.96& 1.3& -0.1& $<$43.1&&&\\
UGC 02982& 11.14& 1.2& 1.3& $<$42.7&&Y&\\
NGC 6621& 11.17& 1.2& 0.7& $<$42.7&&&\\
ESO 550-IG025& 11.44& 1.2& 0.3& $<$42.8&&&\\
IRAS 17578-0400& 11.26& 1.1& 1.4& $<$42.5&&&\\
MCG+02-20-003& 11.03& 1.1& 0.8& $<$42.6&&&\\
PGC 061152& 11.06& 1.1& -0.4& $<$42.8&&&\\
IRAS 05083+2441& 11.22& 1.1& -1.0& $<$43.7&&&\\
NGC 0034& 11.48& 1.1& 0.0& $<$42.8&&&\\
IRAS 23436+5257& 11.50& 1.1& 0.8& $<$42.6&&&\\
IRAS F03359+1523& 11.51& 1.0& 1.3& $<$42.8&&&\\
IRAS F16516-0948& 11.25& 1.0& 1.2& $<$43.1&&&\\
NGC 5010& 10.75& 0.9& 0.3& $<$42.7&&&\\
NGC 7771& 11.31& 0.8& 2.0& $<$42.4&&&\\
MCG-03-04-014& 11.62& 0.8& 0.9& $<$42.6&&&\\
PGC 061675& 11.08& 0.8& 0.1& $<$42.8&&&\\
IRAS F16399-0937& 11.48& 0.8& -0.1& $<$43.2&&&\\
NGC 5331& 11.53& 0.8& -0.6& $<$43.0&&&\\
ESO 593-IG008& 11.86& 0.7& 1.8& $<$42.9&&&\\
NGC 2146& 10.75& 0.7& -1.0& $<$43.0&&&\\
IC 0860& 11.05& 0.7& 0.3& $<$42.3&&&\\
NGC 3110& 11.20& 0.7& 0.1& $<$42.6&&&\\
UGC 02238& 11.29& 0.7& 0.1& $<$42.8&&Y&\\
UGC 04881& 11.64& 0.7& 0.0& $<$43.2&&&\\
NGC 0023& 11.04& 0.6& 1.7& $<$42.5&&&\\
NGC 2623& 11.51& 0.6& 1.3& $<$42.7&&Y&\\
MCG +01-42-008& 11.35& 0.6& 0.8& $<$43.1&&&\\
UGC 08739& 11.02& 0.6& -0.2& $<$42.5&&&\\
IRAS 05442+1732& 11.25& 0.6& 0.4& $<$43.3&&&\\
IRAS F16164-0746& 11.50& 0.6& 0.3& $<$43.2&&&\\
VV 340a& 11.64& 0.6& -0.1& $<$42.2&&&\\
UGC 01845& 11.08& 0.5& 0.9& $<$42.5&&&\\
VV 250a& 11.74& 0.5& 0.8& $<$43.0&&&\\
NGC 5257& 11.37& 0.5& 0.7& $<$42.9&&Y&\\
ESO 507-G070& 11.40& 0.5& 0.1& $<$42.9&&&\\
MCG+07-23-019& 11.54& 0.5& -0.3& $<$43.3&&&\\
NGC 0695& 11.65& 0.5& -0.7& $<$42.6&&&\\
IC 0564& 11.13& 0.4& 1.0& $<$42.8&&&\\
NGC 6090& 11.50& 0.4& -0.2& $<$43.0&&&\\
IRAS F05187-1017& 11.25& 0.3& 1.1& $<$43.1&&&\\
NGC 4194& 10.90& 0.3& 1.0& $<$42.6&&&\\
NGC 1797& 11.00& 0.3& 1.0& $<$42.5&&&\\
UGC 11041& 10.98& 0.3& 0.5& $<$43.0&&&\\
NGC 1614& 12.29& 0.3& 0.4& $<$42.5&&&\\
IRAS 03582+6012& 11.37& 0.3& -0.6& $<$42.6&&Y&\\
II Zw 096& 11.90& 0.2& 1.5& $<$42.3&&&\\
MCG+12-02-001& 11.45& 0.2& 1.4& $<$42.5&&Y&\\
UGC 08387& 11.58& 0.2& 1.0& $<$42.4&&&\\
NGC 0317B& 11.17& 0.2& 0.5& $<$42.6&&&\\
NGC 0992& 11.00& 0.2& 0.3& $<$42.4&&&\\
PGC 014069& 11.14& 0.2& 0.1& $<$42.9&&&\\
NGC 6286& 11.26& 0.2& -1.2& $<$42.6&&&\\
IRAS F06076-2139& 11.61& 0.2& 0.2& $<$43.3&&&\\
NGC 7592& 11.36& 0.1& 1.1& $<$43.0&&&\\
NGC 3221& 10.97& 0.1& 0.4& $<$42.7&&&\\
Mrk 1490& 11.30& 0.1& -0.2& $<$42.8&&&\\
NGC 0828& 11.31& 0.1& -1.9& $<$42.6&&&\\
MCG -02-01-052& 11.45& 0.0& 0.6& $<$41.9&&&\\
PGC 054330& 11.14& 0.0& -0.6& $<$42.4&&&\\
IRAS F12224-0624& 11.20& 0.0& -0.1& $<$43.0&&&\\
NGC 4922& 11.28& -0.1& 2.0& $<$42.8&&Y&\\
NGC 2342& 11.05& -0.1& 0.8& $<$42.7&&&\\
IRAS 05223+1908& 11.57& -0.1& 0.7& $<$42.5&&&\\
NGC 6701& 11.00& -0.1& 0.4& $<$42.3&&&\\
NGC 6670& 11.59& -0.1& 0.4& $<$43.0&&&\\
NGC 5395& 10.68& -0.1& 0.1& $<$42.3&&&\\
IC 4280& 10.98& -0.2& 0.2& $<$42.5&&&\\
UGC 01385& 11.00& -0.2& -0.5& $<$42.7&&&\\
NGC 5936& 10.96& -0.3& -0.5& $<$42.8&&Y&\\
NGC 5653& 10.93& -0.4& 0.6& $<$43.0&&&\\
ESO 602-G025& 11.30& -0.4& 0.4& $<$43.0&&&\\
NGC 6052& 10.88& -0.4& -0.1& $<$42.8&&&\\
VV 283& 11.53& -0.4& -0.2& $<$41&&&\\
IC 2810& 11.05& -0.4& -0.8& $<$42.5&&&\\
IRAS F17138-1017& 11.37& -0.4& -1.2& $<$42.8&&&\\
IRAS F01364-1042& 11.77& -0.4& -0.8& $<$42.9&&&\\
NGC 6907& 10.91& -0.5& 0.1& $<$43.2&&&\\
UGC 03410& 10.93& -0.5& -0.6& $<$43.0&&&\\
MCG+08-18-013& 11.28& -0.5& -1.0& $<$42.9&&&\\
IC 1623A& 11.67& -0.5& -1.1& $<$42.8&&&\\
IRAS F08339+6517& 11.04& -0.6& -0.1& $<$42.6&&&\\
MCG+08-11-002& 11.38& -0.6& -0.4& $<$42.9&&&\\
NGC 7752/3& 10.84& -0.6& -0.6& $<$42.6&&&\\
NGC 2388& 11.11& -0.6& -0.8& $<$42.4&&Y&\\
NGC 6786& 11.27& -0.7& -0.7& $<$42.9&&&\\
UGC 03351& 11.27& -0.8& 0.5& $<$42.6&&&\\
UGC 09913& 12.13& -0.8& 0.3& $<$42.7&&&\\
IRAS F10173+0828& 11.75& -0.8& -0.3& $<$42.5&&&\\
IRAS 20351+2521& 11.56& -0.9& 0.5& $<$42.6&&&\\
NGC 7591& 11.04& -0.9& 0.0& $<$42.6&&&\\
MCG +00-29-023& 11.26& -0.9& -0.7& $<$42.9&&&\\
UGC 12150& 11.29& -0.9& -1.3& $<$42.8&&&\\
IRAS 05129+5128& 11.40& -1.0& -1.9& $<$43.1&&&\\
MCG +02-04-025& 11.66& -1.1& 0.5& $<$43.0&&&\\
NGC 0958& 11.13& -1.1& 0.1& $<$42.7&&&\\
IRAS 21101+5810& 11.72& -1.2& 0.1& $<$43.1&&&\\
MCG+05-06-036& 11.60& -1.2& -0.2& $<$43.2&&&\\
NGC 5734& 10.94& -1.2& -0.9& $<$43.2&&&\\
UGC 02369& 11.70& -1.3& 1.0& $<$41.8&&&\\
IC 0214& 11.39& -1.4& -0.5& $<$42.2&&&\\
III Zw 035& 11.60& -1.6& -0.3& $<$43.0&&&\\
IRAS F10565+2448& 11.99& -1.7& -1.0& $<$43.3&&&\\
VV 705& 12.33& -1.8& 0.6& $<$42.7&&Y&\\
PGC 070417& 11.31& -2.0& -1.3& $<$42.9&&&\\
MCG -01-60-022& 11.16& -2.1& -2.7& $<$42.9&&&\\

\footnotetext[1]{IR luminosity ($L_{8-1000 \; \mu m }$$>$$10^{11} L_{\sun}$) based on SED fitting \citep{Casey:12:1595} using data from IRAS.}
\footnotetext[2]{BAT Signal to Noise Ratio (SNR) defined as the background-subtracted source count rate divided by local background standard deviation.}
\footnotetext[3]{BAT luminosity and 1$\sigma$ error.  Lower limits were calculated at 3$\sigma$, using an X-ray power law of $\Gamma$=1.9, and Galactic extinction, consistent with the mean 14-195 power law for Seyfert 2s in the BAT sample \citep{Winter:09:1322}. }
\footnotetext[4]{Ultra hard X-ray hardness flux ratio ($HR_{UX}$=14-195 keV/2-10 keV).}
\footnotetext[5]{Presence of NeV from \citet{Petric:11:28}. }
\footnotetext[6]{2-10 keV references where G08=\citet{Greenhill:08:L13}, I11=\citet{Iwasawa:11:106}, L10=\citet{Lehmer:10:559}, P11=\citet{Pereira-Santaella:11:93}, W09=\citet{Winter:09:1322}}
\footnotetext[7]{? indicates no available high quality 2-10 keV measurement.}
\end{longtable}
\end{center}

\end{document}